\documentclass
[superscriptaddress,secnumarabic,amssymb,amsmath,nobibnotes,aps,prd,showkeys,showpacs,nofootinbib,twocolumn]{revtex4}%
\usepackage{graphicx}
\usepackage{epsf}
\usepackage{bm}
\usepackage{amsmath}
\usepackage{amsfonts}
\usepackage{amssymb}
\usepackage{epstopdf}%
\providecommand{\U}[1]{\protect\rule{.1in}{.1in}}

\newcommand{\be}{\begin{equation}}
\newcommand{\ee}{\end{equation}}

\newcommand{\mincir}{\raise
-3.truept\hbox{\rlap{\hbox{$\sim$}}\raise4.truept\hbox{$<$}\ }}
\newcommand{\magcir}{\raise
-3.truept\hbox{\rlap{\hbox{$\sim$}}\raise4.truept\hbox{$>$}\ }}

\begin{document}
\title{On the Hojman conservation quantities in Cosmology}
\author{A. Paliathanasis}
\email{anpaliat@phys.uoa.gr}
\affiliation{Instituto de Ciencias F\'{\i}sicas y Matem\'{a}ticas, Universidad Austral de
Chile, Valdivia, Chile.}
\author{P.G.L. Leach}
\email{leach.peter@ucy.ac.cy}
\affiliation{Department of Mathematics and Statistics,University of Cyprus, Lefkosia 1678, Cyprus}
\affiliation{Department of Mathematics and Institute of Systems Science, Research and
Postgraduate Support, Durban University of Technology, PO Box 1334, Durban
4000, Republic of South Africa}
\affiliation{School of Mathematics, Statistics and Computer Science, University of
KwaZulu-Natal, Private Bag X54001, Durban 4000, Republic of South Africa.}
\author{S. Capozziello}
\email{capozziello@na.infn.it}
\affiliation{Dipartimento di Fisica, Universita' di Napoli Federico II, Complesso
Universitario di Monte S. Angelo, Via Cinthia, 9, I-80126 Naples, Italy.}
\affiliation{Istituto Nazionale di Fisica Nucleare (INFN) Sez. di Napoli, Complesso
Universitario di Monte S. Angelo, Via Cinthia, 9, I-80126 Naples, Italy.}
\affiliation{Gran Sasso Science Institue (INFN), Viale F. Crispi 7, I-67100, L' Aquila, Italy.}
\affiliation{Tomsk State Pedagogical University, ul. Kievskaya, 60, 634061 Tomsk, Russia.}

\begin{abstract}
We discuss the application of the Hojman's Symmetry Approach for the
determination of conservation laws in Cosmology, which has been recently
applied by various authors in different cosmological models. We show that
Hojman's method for regular Hamiltonian systems, where the Hamiltonian
function is one of the involved equations of the system, is equivalent to the
application of Noether's Theorem for generalized transformations. That means
that for minimally-coupled scalar field cosmology or other modified theories
which are conformally related with scalar-field cosmology, like $f(R)$
gravity, the application of Hojman's method provide us with the same results
with that of Noether's theorem. Moreover we study the special Ansatz.
$\phi\left(  t\right)  =\phi\left(  a\left(  t\right)  \right)  $, which has
been introduced for a minimally-coupled scalar field, and we study the Lie and
Noether point symmetries for the reduced equation. We show that under this
Ansatz, the unknown function of the model cannot be constrained by the
requirement of the existence of a conservation law and that the Hojman
conservation quantity which arises for the reduced equation is nothing more
than the functional form of Noetherian conservation laws for the
free particle. On the other hand, for $f(T)$ teleparallel gravity, it is not
the existence of Hojman's conservation laws which provide us with the special
function form of $f(T)$ functions, but the requirement that the reduced
second-order differential equation admits a Jacobi Last multiplier, while the
new conservation law is nothing else that the Hamiltonian function of the
reduced equation.

\end{abstract}
\maketitle
\date{\today}

The source for the late-time cosmic acceleration
\cite{Teg,Kowal,Komatsu,planck} has been attributed to an unidentified type of
matter-energy with a negative parameter in the equation of state, the dark
energy. The cosmological constant, $\Lambda$, is the simplest candidate for
dark energy with a parameter in the equation of state, $w_{\Lambda}=-1$.
However, the cosmological constant suffers from two major problems. They are
the fine tuning and the coincidence problems \cite{Padmanabhan1,Weinberg1}.
Consequently other dark energy candidates have been introduced, such as
cosmologies with time-varying $\Lambda\left(  t\right)  $, quintessence,
Chaplygin gas, matter creation, $f\left(  R\right)  $ gravity and many others.
This status of art imposes a discrimination among the various cosmological
models and the search for new approaches to find out exact solutions to be
matched with data. In particular, one of the issue is that cosmological models
should come out from some first principles in order to be related to some
fundamental theory.

Below we restrict our consideration to a spatially flat
Friedmann--Robertson--Walker (FRW) spacetime with line element%
\begin{equation}
ds^{2}=-dt^{2}+a^{2}\left(  t\right)  \left(  \mbox{\rm d}x^{2}%
+\mbox{\rm d}y^{2}+\mbox{\rm d}z^{2}\right)  . \label{let.01}%
\end{equation}

Let us start our considerations from cosmological solutions derived from
General Relativity (GR). Standard GR provides us with a set of second-order
differential equations. Consider a Riemannian manifold $\mathcal{M}^{4}%
,~$induced with a metric tensor $g_{ij},$ and GR with cosmological constant,
$\Lambda$. The Action integral of the field equations is as follows:
\begin{equation}
S=\int dx^{4}\sqrt{-g}\left(  R-2\Lambda\right)  , \label{let.02}%
\end{equation}
where $R$ is the Ricci scalar of $\mathcal{M}^{4}$. For the spacetime,
(\ref{let.01}), the Ricci scalar is
\begin{equation}
R=6\left[  \frac{\ddot{a}}{a}+\left(  \frac{\dot{a}}{a}\right)  ^{2}\right]  ,
\end{equation}
and from the variational principle for the Action integral, (\ref{let.02}), we
have the set of differential equations,
\begin{equation}
-3a\dot{a}^{2}+2a^{3}\Lambda=0 \label{let.03}%
\end{equation}
and
\begin{equation}
\ddot{a}+\frac{1}{2a}\dot{a}^{2}-a\Lambda=0, \label{let.04}%
\end{equation}
where $\dot{a}=da/dt$. \ It is well known that the solution of the system
(\ref{let.03}, \ref{let.04}) is
\begin{equation}
a\left(  t\right)  =a_{0}\exp\left[  \sqrt{\frac{2\Lambda}{3}}t\right]
\end{equation}
which is the de Sitter solution. In terms of dynamics, the system
(\ref{let.03}, \ref{let.04}) is that of a one-dimensional hyperbolic
oscillator for which equation (\ref{let.03}) can be interpreted as the
Hamiltonian constraint on the oscillator. There are two ways to observe this.
The first way is to apply the \textquotedblleft coordinate
transformation\textquotedblright\ $a\left(  t\right)  =r\left(  t\right)
^{\frac{2}{3}}$. When we use this transformation, the system, (\ref{let.03},
\ref{let.04}), reduces to the simplest form,
\begin{equation}
-\frac{1}{2}\dot{r}^{2}+\frac{\omega^{2}}{2}r^{2}=0~,~~\ddot{r}-\omega
^{2}r=0~~\mbox{\rm with}~~\omega^{2}=\frac{3}{2}\Lambda. \label{let.05}%
\end{equation}
The second way is to study the point symmetries. We consider the Noether point
symmetries of the Lagrangian of the field equations. The Lagrangian which
follows from (\ref{let.02}) is
\begin{equation}
L\left(  a,\dot{a}\right)  =3a\dot{a}^{2}+2\Lambda a^{3}%
\end{equation}
and we can easily see that this admits five Noether point symmetries.
Moreover, because $L\left(  a,\dot{a}\right)  $ admits the Noether algebra of
maximal dimension, this indicates that $L\left(  a,\dot{a}\right)  $ can
describe two systems: the one-dimensional free particle and the
one-dimensional linear equation\footnote{They are mathematically equivalent
under a point transformation, but we prefer to maintain a physical
distinction.} \cite{Leach81}. However, as the potential in $L\left(  a,\dot
{a}\right)  $ is not constant, we can say that the dynamical system is the
one-dimensional linear system and, specifically, that of the hyperbolic oscillator, when $\Lambda$ is a positive constant.

Consider now as a candidate for dark energy a quintessence scalar field with
action%
\begin{equation}
S_{\phi}=\int dx^{4}\sqrt{-g}\left(  -\frac{1}{2}g_{\mu\nu}\phi^{;\mu}%
\phi^{;\nu}+V\left(  \phi\right)  \right)  . \label{let.06}%
\end{equation}

Consequently, for a model with a scalar field, the Action integral of the
field equations is
\[
S=\int dx^{4}\sqrt{-g}R+S_{\phi},
\]
and for the FRW spacetime, (\ref{let.01}), the field equations comprise the
set of differential equations,
\begin{equation}
-3a\dot{a}^{2}+\frac{1}{2}a^{3}\dot{\phi}^{2}+V\left(  \phi\right)  =0,
\label{let.07}%
\end{equation}
and
\begin{equation}
\ddot{a}+\frac{1}{2a}\dot{a}^{2}+\frac{a}{2}\dot{\phi}^{2}-aV\left(
\phi\right)  =0, \label{let.08}%
\end{equation}
where the scalar field, $\phi\left(  t\right)  $, satisfies the equation%
\begin{equation}
\ddot{\phi}+\frac{3}{a}\dot{a}\dot{\phi}+V\left(  \phi\right)  _{,\phi}=0.
\label{let.09}%
\end{equation}

Similarly equation (\ref{let.07}) can be viewed as the Hamiltonian
constraint\footnote{We note that in a lapse time $\mbox{\rm d}t=N\left(
\tau\right)  \mbox{\rm d}\tau$ of (\ref{let.01}), equations (\ref{let.03}) and
(\ref{let.07}) arise from the variation of the variable $N\left(  \tau\right)
$ and are restricted to that form when we consider $N\left(  \tau\right)
=1$.} of equations (\ref{let.08}) and (\ref{let.09}). In order that we can
study the (Liouville) integrability of the Hamiltonian system (\ref{let.07}%
)-(\ref{let.09}), we have to study the existence of conservation laws. As the
system has dimension two, being the configuration space ${\mathbb{Q}}%
\equiv\left\{  a,\phi\right\}  $, and (\ref{let.07}) can be seen as a
conservation law, we have to determine the exact form of the potential
$V\left(  \phi\right)  $ in which the system admits additional conservation
laws. We remark that Liouville integrability means that the field equations
can be reduced to quadratures from which we can seek for determining the
solution of the scalar factor, $a\left(  t\right)  $, in a closed form. The
most common method to determine conservation laws for Hamiltonian systems is
the well-known Noether Symmetry Approach. This has lead many researchers to
the study of the Noether symmetries for dynamical system (\ref{let.07}%
)-(\ref{let.09}) (see for example \cite{Sf1,Sf2,Sf6,Sf7} and reference therein).

Recently, in \cite{Cap1}, it has been proposed that the unknown potential of
the scalar field be constrained by the existence of Hojman conserved
quantities. The results have been applied also to the cosmological scenario
with a nonminimally coupled scalar-field model \cite{Cap2}, in $f\left(
R\right)  $-gravity in the metric formalism \cite{wei1} and in $f\left(
T\right)  $-gravity \cite{wei2}. In what follows we follow the notation by
\cite{Cap1,Cap2}.

Hojman proved that, if a system of second-order differential equations,
\begin{equation}
\ddot{q}^{i}=F^{i}\left(  t,q^{k},\dot{q}^{k}\right)  , \label{let.10}%
\end{equation}
is invariant under the transformation,
\begin{equation}
q^{\prime i}=q^{i}+X^{i}\left(  t,q^{k},\dot{q}^{k}\right)  , \label{let.11}%
\end{equation}
then the quantity $Q\left(  t,q^{i},\dot{q}^{i}\right)  $ with form
\begin{equation}
Q=\frac{\partial X^{i}}{\partial q^{i}}+\frac{\partial}{\partial\dot{q}^{i}%
}\left(  \frac{dX^{i}}{dt}\right)  \label{let.12}%
\end{equation}
for
\begin{equation}
\frac{\partial F^{i}}{\partial\dot{q}^{i}}=0, \label{let.12a}%
\end{equation}
or
\begin{equation}
Q=\frac{1}{\gamma}\frac{\partial\left(  \gamma X^{i}\right)  }{\partial q^{i}%
}+\frac{\partial}{\partial\dot{q}^{i}}\left(  \frac{dX^{i}}{dt}\right)  ~
\label{let.13}%
\end{equation}
for
\begin{equation}
\frac{\partial F^{i}}{\partial\dot{q}^{i}}=-\frac{d}{dt}\ln\gamma,
\label{let.13a}%
\end{equation}
is a conservation law of (\ref{let.10}), i.e. $\frac{dQ}{dt}=0$, where
$\gamma=\gamma\left(  q^{k}\right)  ~$\cite{Hojman}. \ We remark that the
differential equation (\ref{let.10}) is invariant under the action of
(\ref{let.11}) and this means that the following condition holds%
\begin{equation}
\ddot{X}^{i}-\frac{\partial F^{i}}{\partial q^{j}}X^{j}-\frac{\partial F^{i}%
}{\partial\dot{q}^{j}}\dot{X}^{j}=0. \label{let.14}%
\end{equation}
Furthermore, conditions (\ref{let.12a}) or (\ref{let.13a}) are necessary to
hold in order the conservation law to exist.

In order to simplify the problem the authors in \cite{Cap1} reduced the
two-dimensional dynamical system $\left\{  a,\phi\right\}  $ of the field
equations to a one-dimensional system in $\left\{  a\right\}  $ by selecting
$\phi\left(  t\right)  =\phi\left(  a\left(  t\right)  \right)  $. \ When this
is used, equation (\ref{let.08}), with the use of (\ref{let.07}), becomes%
\begin{equation}
\ddot{x}+f\left(  x\right)  \dot{x}^{2}=0, \label{let.15}%
\end{equation}
where $x=\ln a,$ and $f\left(  x\right)  =\frac{1}{2}\left(  \frac
{d\phi\left(  x\right)  }{dx}\right)  ^{2}$. Moreover, from equation
(\ref{let.09}) and (\ref{let.07}), we have%
\begin{equation}
\left(  \ln V\left(  \phi\right)  \right)  _{,\phi}=\frac{f\left(  x\right)
\phi_{,x}-\phi_{,xx}-3\phi_{,x}}{3-\frac{1}{2}\left(  \phi_{,x}\right)  ^{2}}.
\label{let.16}%
\end{equation}

In \cite{Cap1} the authors substituted $F\left(  t,x,\dot{x}\right)
=-f\left(  x\right)  \dot{x}^{2}$ into (\ref{let.14}) and constrained the
function $f\left(  x\right)  $ in order that equation (\ref{let.15}) admit
Hojman conservation laws. Furthermore from (\ref{let.16}) they determined the
potential $V\left(  \phi\right)  $ and found closed-form cosmological solutions.

We continue by studying the Lie point symmetries of the second-order
differential equation (\ref{let.15}). We consider a point transformation in
the space $\left\{  t,x\right\}  $ and we find that equation (\ref{let.15}) is
invariant under an eight-dimensional Lie algebra of point transformations,
automatically $sl\left(  3,R\right)  $, for arbitrary functional form of
$f\left(  x\right)  $. Therefore the Lie theorem holds, i.e., it means that
there exists a \textquotedblleft coordinate\textquotedblright\ transformation
$\left\{  t,x\right\}  \rightarrow\left\{  \tau,y\right\}  $, whereby equation
(\ref{let.15}) is reduced to that of the free particle, $y^{\prime\prime}%
=0~$\cite{StephaniBook}. By using the Lie symmetries, we have that the
transformation is
\begin{equation}
t=t~~,~~y=\int\exp\left(  \int f\left(  x\right)  dx\right)  dx \label{let.17}%
\end{equation}
whence (\ref{let.16}) becomes $\ddot{y}=0$. Hence $y\left(  t\right)
=y_{1}t+y_{0}$, $y_{1},y_{0}\in%
\mathbb{R}
$, that is $\int\exp\left(  \int f\left(  x\right)  dx\right)  dx=y_{1}%
t+y_{0}$ which is the solution for the scale factor $a\left(  t\right)
=\exp[x\left(  t\right)  ]$ for an arbitrary function, $f\left(  x\right)  $.
Moreover the potential is constrained by equation (\ref{let.16}).

The function $L_{f}\left(  y,\dot{y}\right)  =\frac{1}{2}\dot{y}^{2}$, is one
of the possible Lagrangians of the free particle. It is straightforward to see
that $L_{f}$ admits five Noetherian point symmetries, the vector fields%
\begin{equation}
Z_{1}=\partial_{t}~,~Z_{2}=\partial_{y}~,~Z_{3}=t\partial_{y} \label{let.18}%
\end{equation}%
\begin{equation}
Z_{4}=2t\partial_{t}+y\partial_{y}~,~Z_{5}=t^{2}\partial_{t}+ty\partial_{y}
\label{let.18b}%
\end{equation}
for which the corresponding Noetherian conservation laws are%
\begin{align}
Z_{1}  &  :\mathcal{H}_{f}=\frac{1}{2}\dot{y}^{2}~,~Z_{2}:I_{p}=\dot
{y}~,~Z_{3}:I_{p}^{^{\prime}}=t\dot{y}-y\label{let.19}\\
Z_{4}  &  :I_{s}=2t\mathcal{H}_{f}-y\dot{y}~,~Z_{5}:I_{s}^{\prime}%
=t^{2}\mathcal{H}_{f}-ty\dot{y}+\frac{1}{2}y^{2}. \label{let.20}%
\end{align}

We remark that the conservation laws, $\mathcal{H}_{f}$, and $I_{p}$, are the
well-known conservation of energy and momentum, in particular it holds that
$\mathcal{H}_{f}=\frac{1}{2}\left(  I_{p}\right)  ^{2}$. Furthermore, under
the coordinate transformation (\ref{let.17}), the momentum, $I_{p}$, becomes
$I_{p}=\exp\left(  \int f\left(  x\right)  dx\right)  \dot{x}$.

We continue with the determination of the Hojman conservation laws for
equation (\ref{let.15}). Without loss of generality, we study the Hojman
conservation laws of $\ddot{y}=0$. For this equation, condition (\ref{let.14})
becomes \ $\ddot{X}=0$, that is,%
\begin{equation}
X_{,tt}+2\dot{y}X_{,ty}+\dot{y}^{2}X_{,yy}=0. \label{let.21}%
\end{equation}
Hence
\begin{equation}
X\left(  x,y,\dot{y}\right)  =X_{1}\left(  t\dot{y}-y,\dot{y}\right)
+tX_{2}\left(  t\dot{y}-y,\dot{y}\right)  ,
\end{equation}
or%
\[
X\left(  x,y,\dot{y}\right)  =X_{1}\left(  I_{p},I_{p}^{\prime}\right)
+tX_{2}\left(  I_{p},I_{p}^{\prime}\right)  .
\]
It is easy to see that, when $\frac{\partial X}{\partial\dot{y}}=0$, we have
the Lie~ symmetries $Z_{2},Z_{3}~$and $Y_{L}=y\partial_{y}$. Therefore from
(\ref{let.12}) we have the general Hojman conservation law%
\begin{equation}
Q\left(  t,y,\dot{y}\right)  =Q(I_{p}^{\prime},I_{p})=\frac{\partial}{\partial
I_{p}^{\prime}}X_{1}(I_{p}^{\prime},I_{p}), \label{let.22}%
\end{equation}

Consequently, for equation (\ref{let.15}), the Hojman conservation law has the
following form
\begin{equation}
Q\left(  t,x,\dot{x}\right)  =Q\left(  I_{p}\left(  t,x,\dot{x}\right)
,I_{p}^{^{\prime}}\left(  t,x,\dot{x}\right)  \right)  . \label{let.23}%
\end{equation}
where now%
\begin{equation}
I_{p}=\exp\left(  \int f\left(  x\right)  dx\right)  \dot{x}%
\end{equation}%
\begin{equation}
I_{p}^{^{\prime}}=t\exp\left(  \int f\left(  x\right)  dx\right)  \dot{x}%
-\int\exp\left(  \int f\left(  x\right)  dx\right)  dx
\end{equation}
We have proved that equation (\ref{let.15}) admits conservation laws for an
arbitrary function $f\left(  x\right)  ,$ and that the Hojman conservation law
$Q\left(  I_{p}\right)  $ for that case is a functional form of Noetherian
conservation laws for the free particle.

We conclude that, under the Ansatz. $\phi\left(  t\right)  =\phi\left(
a\left(  t\right)  \right)  $, the reduced field equation (\ref{let.15}) is
integrable for arbitrary function $f\left(  x\right)  $, i.e. Eq.
(\ref{let.15}) admits conservation laws for any $f\left(  x\right)  $. On the
other hand, the Ansatz. $\phi=\phi\left(  a\right)  $, reduces the field
equation in a one-dimensional autonomous dynamical system, which is
integrable. Moreover the Hojman conservation quantities for equation
(\ref{let.15}) are functions of the Noetherian conservation laws of the free
particle. Therefore there are no difference among the two methods at that
level, that is, the claim in \cite{Cap1} that the class of scalar field
potentials $V\left(  \phi\right)  $ is constrained by condition (\ref{let.14})
for equation, (\ref{let.15}) means that Hojman conservation laws exists when
Noetherian conservation laws exists.

Furthermore we remark that the solutions which have been found in \cite{Cap1}
are special solutions and are not the general solutions of the field equations
in the sense that they hold when $\phi=\phi\left(  a\right)  $ and they are
restricted to the case for which there exists an inverse function
$t=a^{-1}\left(  t\right)  $. This can be seen easily for the closed-form
solution are given for the exponential potential $V\left(  \phi\right)
=V_{0}e^{\lambda\phi}$ in \cite{Cap1}. The solution found there is a power
law. However, the solution for that model is more general and can be found in
\cite{Russo}.

As far as concerns the application of Hojman's conservation quantities in
nonminimally coupled scalar field cosmology \cite{Cap2}, or in $f\left(
R\right)  $ gravity in the metric formalism \cite{wei1} the situation is
similar with above. Note that $f\left(  R\right)  $ gravity in the metric
formalism can be seen as a Brans-Dicke-like model with vanishing Brans-Dicke
parameter, i.e. the O' Hanlon gravity \cite{Hanlon}. These theories are
related under conformal transformations and are equivalent with the minimally
coupled scalar field \cite{Rept}. Moreover when there is no other matter
source, then the solution of the field equations holds either in the Einstein
or in the Jordan frame \cite{MtGRG}.

Because of that, the same Ansatz for the application of Hojman's method in
scalar tensor theories can be used, i.e. the field \textquotedblleft$\phi
$\textquotedblright\ to be a function of the scale factor. Hence the field
equations reduced to a second-order ordinary differential equation of the form
of (\ref{let.15}). Therefore the above analysis and comments hold.

In the case of $f\left(  T\right)  $ gravity the situation is different since
the field equations are a singular one-dimensional dynamical system.

Following \cite{wei2}, we take into account a spatially flat FRW spacetime
(\ref{let.01}) and the basis $e^{\alpha}\left(  x^{\mu}\right)  =h_{\mu
}^{\alpha}dx^{\mu}$, \ such as, $g_{\mu\nu}=\eta_{\alpha\beta}e^{\alpha
}e^{\beta}=\eta_{a\beta}h_{\mu}^{\alpha}h_{\mu}^{\beta},~$where%
\begin{equation}
h_{\mu}^{i}=diag\left(  1,e^{x\left(  t\right)  },e^{x\left(  t\right)
},e^{x\left(  t\right)  }\right)  . \label{let.25}%
\end{equation}
$f\left(  T\right)  $ gravity is a straightforward extension of Teleparallel
GR (TEGR), such as $f\left(  R\right)  $gravity of GR. The Action integral in
$f\left(  T\right)  $ gravity with a matter term is%
\begin{equation}
A_{T}=\int dx^{4}\left\vert e\right\vert f\left(  T\right)  +\int
dx^{4}\left\vert e\right\vert L_{m}, \label{let.26}%
\end{equation}
where~$L_{m}$ is the Lagrangian of the matter term, and $T$ is a function of
the Torsion scalar (for details see \cite{ft1,manos}). For the basis
(\ref{let.25}) from the action integral \ (\ref{let.26}), for a perfect fluid
with constant equation of state parameter $p=w\rho$, we have that
$T=-6H^{2},~H=\dot{x}\left(  t\right)  $, the field equations are:%
\begin{equation}
12f_{,T}H+f=\rho\label{let.27}%
\end{equation}%
\begin{equation}
48H^{2}f_{,TT}\dot{H}-f_{,T}\left(  12H^{2}+4\dot{H}\right)  -f=p
\label{let.28}%
\end{equation}
and the conservation law for the fluid, $\dot{\rho}+3\left(  1+w\right)  \rho
H=0,~$which gives $\rho=\rho_{m0}e^{-3\left(  1+w\right)  x}$.~$\ $The field
equations (\ref{let.27}), (\ref{let.28}) form a singular one-dimensional
Hamiltonian system since $\det\left\vert \frac{\partial L}{\partial x^{i}%
}\right\vert = 0$. Furthermore equation (\ref{let.28}) is a second-order
differential equation of $x\left(  t\right)  $, while (\ref{let.27}) is a
first order equation of $x\left(  t\right)  $. Typically, the system is
integrable however it is not always possible to find the solution in
closed-form. A classical analogue is the closed form solution of the
one-dimensional Newtonian system, $E=\frac{1}{2}\dot{y}^{2}+V\left(  y\right)
,$which admits closed-form solutions for functions $V\left(  x\right)  $,
where the equations of motion admit Lie symmetries. For the field equations
(\ref{let.27}), (\ref{let.28}), the unknown function is defined by the
Lagrange multiplier $T=-6H^{2}$, where, again, the existence of point
symmetries provides closed-form solutions. With the use of (\ref{let.27}) and
the Lagrange multiplier $T$, equation (\ref{let.28}) can be written in the
following form,%
\begin{equation}
\ddot{x}=F\left(  \dot{x}\right)  , \label{let.30}%
\end{equation}
where $F\left(  \dot{x}\right)  =F\left(  T\right)  =F\left(  f\left(
T\right)  \right)  $. In order to exist a Hojman conservation law for the
second-order differential equation (\ref{let.30}), condition (\ref{let.12a})
or (\ref{let.13a}) have to hold. In other words, $F\left(  \dot{x}\right)
=-F_{1}\dot{x}^{2}-F_{0}$. This means that equation (\ref{let.30}) is
\begin{equation}
\ddot{x}+F_{1}\dot{x}^{2}+F_{0}=0 \label{let.31}%
\end{equation}
where the general solution is
\begin{equation}
x\left(  t\right)  =\frac{1}{2F_{1}}\ln\left[  \frac{F_{1}}{F_{0}}\left(
x_{1}\cos\left(  \sqrt{F_{1}F_{0}}t\right)  -x_{2}\sin\left(  \sqrt{F_{1}%
F_{0}}t\right)  \right)  ^{2}\right]  . \label{let.32}%
\end{equation}

However, equation (\ref{let.31}) admits eight Lie point symmetries, i.e. the
$sl\left(  3,R\right)  $, and according to the Lie theorem, it is equivalent
to the free particle, $y^{\prime\prime}=0$. Moreover it is easy to see that
equation (\ref{let.31}) follows from the Lagrangian
\begin{equation}
L\left(  x,\dot{x}\right)  =\frac{1}{2}e^{2F_{1}x}\dot{x}^{2}-F_{0}e^{2F_{1}%
x}, \label{let.33}%
\end{equation}
and, since the latter is autonomous, admits as Noetherian conservation law the
Hamiltonian function. In the case where $F_{0}=0$, (in the notation of
\cite{wei2}, $c=0$), Lagrangian (\ref{let.33}) is that of the free particle
and also the momentum is another time-independent Noetherian conservation law.
That is the conservations laws in \cite{wei2} are not conservation laws which
follows from the Hojman's formula but from Noetherian symmetries in the same
way we discussed above.

However condition (\ref{let.13a}) for equation (\ref{let.30}) is equivalent
with the existence of a Jacobi Last multiplier for equation (\ref{let.30}).
The existence of a Jacobi Last multiplier for one-dimensional second order
differential equations of the form of (\ref{let.10}) is equivalent to the
existence of a Lagrangian \cite{NucciT,Partha}, and, in the simplest case that
the dynamical system is autonomous, Hojman conservation laws are equivalent to
the Noetherian conservation laws. Finally, the functional forms of $f\left(
T\right)  $, which follow from the solution of the system%
\begin{equation}
F\left(  \dot{x}\right)  =-F_{1}\dot{x}^{2}-F_{0},
\end{equation}
are not arising from the existent of a Hojman conservation law, but from the
existence of a Jacobi Last multiplier for the reduced equation (\ref{let.30})
and the conservation laws are Noetherian conservation laws, the well known
conservation laws of the Hamiltonian or of the momentum for the
\textquotedblleft oscillator\textquotedblright\ or the free particle.

In conclusion, for regular dynamical systems, such as scalar-field cosmology,
we consider the derivation of Hojman conservation laws for the field equations
(\ref{let.08}, \ref{let.09}) without the Ansatz. $\phi\left(  t\right)
=\phi\left(  a\left(  t\right)  \right)  $. Hence we have to determine the
group of invariant transformations of the system (\ref{let.08}),
(\ref{let.09}). Moreover, at the same time the Hamiltonian function
$\mathcal{H}=0,~$i.e. equation (\ref{let.08}), should be also invariant, which
means that~the following condition
\begin{equation}
X^{\left[  1\right]  }\left(  \mathcal{H}\right)  =\lambda\mathcal{H},
\label{let.24}%
\end{equation}
has to hold, where $X^{\left[  1\right]  }$ is the first prolongation of $X$
and $\lambda$ is an arbitrary function. However, equation (\ref{let.24}) is
nothing else than the Noether condition in Hamiltonian formalism for
generalized symmetries (see \cite{Sarlet}). In the case of point and contact
transformations, this is a well known result \cite{LeachHojman96,LeachFlessas}%
. Therefore the determination of Hojman conservation quantities in
cosmological models, which arise from a Lagrange function, it is equivalent
with the application of Noether's Theorem. That holds for all the physical
systems where the Hamiltonian function is one of the equations involved.

When the field equations are a singular dynamical system, as in the case of
$f\left(  T\right)  $ gravity, the functional forms of $f\left(  T\right)  $,
in which the field equations admit a closed-form solution, follows from the
existence of a Jacobi Last multiplier, and the conservation laws are again Noetherian.

\begin{acknowledgments}
AP acknowledges financial support of FONDECYT grant no. 3160121 and thanks
Prof. PGL Leach, Sivie Govinder and the Department of Mathematics of the
University of Cyprus for the hospitality provided while part of this work
carried out. SC acknowledges the support of INFN (\textit{iniziative
specifiche} QGSKY and TEONGRAV) and thanks the TSPU for being awarded as
Honorary Professor.
\end{acknowledgments}

\end{document}